\documentclass[aps,prl,twocolumn,superscriptaddress,longbibliography]{revtex4-2}
\usepackage[utf8]{inputenc}
\usepackage[T1]{fontenc}

\usepackage[english]{babel}
\usepackage{graphicx}  % needed for figures
\usepackage{dcolumn}   % needed for some tables
\usepackage{bm}        % for math
\usepackage{amssymb}   % for math
\usepackage{amsmath, latexsym}
\usepackage{units}
\usepackage[dvipsnames]{xcolor}
\usepackage[%
colorlinks=true,
urlcolor=RoyalBlue,
linkcolor=RoyalBlue,
citecolor=RoyalBlue,
]{hyperref}
\usepackage{etoolbox}
\usepackage{xfrac}
\usepackage{orcidlink}

\usepackage{sidecap}
\sidecaptionvpos{figure}{t}

\usepackage[type1]{libertine}                                        % Linux Libertine für zweispaltige Texte
\usepackage{textcomp}% Required to get special symbols
\usepackage[scaled=.85]{beramono}% Typewriter font
\usepackage[libertine,cmintegrals,cmbraces,vvarbb,slantedGreek]{newtxmath}
\usepackage[scr=boondoxo]{mathalfa}% Extra math symbols
\usepackage{bm}% Extra bold faces
\usepackage[lf]{carlito}

\makeatletter
\DeclareRobustCommand{\cev}[1]{%
  \mathpalette\do@cev{#1}%
}
\newcommand{\do@cev}[2]{%
  \fix@cev{#1}{+}%
  \reflectbox{$\m@th#1\vec{\reflectbox{$\fix@cev{#1}{-}\m@th#1#2\fix@cev{#1}{+}$}}$}%
  \fix@cev{#1}{-}%
}
\newcommand{\fix@cev}[2]{%
  \ifx#1\displaystyle
    \mkern#23mu
  \else
    \ifx#1\textstyle
      \mkern#23mu
    \else
      \ifx#1\scriptstyle
        \mkern#22mu
      \else
        \mkern#22mu
      \fi
    \fi
  \fi
}
\makeatother

\usepackage{vmargin}
\setpapersize{A4} \oddsidemargin20mm \textwidth170mm \topmargin10mm
\newcommand{\figref}[2][]{\def\a{#1}\def\e{}Fig.\ \ref{#2}\if\a\e\else (\a)\fi}

\newcommand{\panelcaption}[1]{(#1)}
\newcommand{\panelsubcaption}[1]{(#1)}

\begin{document}

\title{Enhancing the Energy Resolution in Scanning Tunneling Microscopy: from dynamical Coulomb blockade to cavity quantum electrodynamics}

\author{Xianzhe Zeng}
\affiliation{Max-Planck-Institut f\"ur Festk\"orperforschung, Heisenbergstraße 1, 70569 Stuttgart, Germany}
\author{Janis Siebrecht}
\affiliation{Max-Planck-Institut f\"ur Festk\"orperforschung, Heisenbergstraße 1, 70569 Stuttgart, Germany}
\author{Haonan Huang\,\orcidlink{0000-0003-4059-0351}}
\affiliation{Max-Planck-Institut f\"ur Festk\"orperforschung, Heisenbergstraße 1, 70569 Stuttgart, Germany}
\author{Sujoy Karan\,\orcidlink{0000-0002-1319-7733}}
\affiliation{Max-Planck-Institut f\"ur Festk\"orperforschung, Heisenbergstraße 1, 70569 Stuttgart, Germany}
\author{Joachim Ankerhold\,\orcidlink{0000-0002-6510-659X}}
\affiliation{Institute for Complex Quantum Systems and IQST, Universität Ulm, Albert-Einstein-Allee 11, 89069 Ulm, Germany}
\author{Klaus Kern\,\orcidlink{0000-0002-1785-7874}}
\affiliation{Max-Planck-Institut f\"ur Festk\"orperforschung, Heisenbergstraße 1, 70569 Stuttgart, Germany}
\affiliation{Institut de Physique, Ecole Polytechnique F{\'e}d{\'e}rale de Lausanne, 1015 Lausanne, Switzerland}
\author{Christian R. Ast\,\orcidlink{0000-0002-7469-1188}}
\email[Corresponding author; electronic address:\ ]{c.ast@fkf.mpg.de}
\affiliation{Max-Planck-Institut f\"ur Festk\"orperforschung, Heisenbergstraße 1, 70569 Stuttgart, Germany}

\date{\today}

\begin{abstract}
    Scanning tunneling microscopy and spectroscopy have become indispensable tools for probing condensed matter at atomic length scales, yet achieving ultimate energy resolution remains a persistent challenge. At mK temperatures, the dynamical Coulomb blockade regime fundamentally limits spectroscopic precision through energy exchange between tunneling electrons and the electromagnetic environment. Here, we demonstrate that combining local electromagnetic shielding with low-pass filtering directly at the cryogenic scan head improves the energy resolution by nearly an order of magnitude, reaching benchmark values as low as 3.7\,$\upmu$eV at 10\,mK. We attribute this enhancement to efficient suppression of high-frequency radiation and capacitive shunting of the tunnel junction. Remarkably, this improved sensitivity reveals that the Josephson current couples to electromagnetic cavity modes of the centimeter-scale scan head, establishing a direct connection between atomic-scale tunneling processes and macroscopic cavity quantum electrodynamics. These advances open pathways for exploring ultra-low-energy phenomena with unprecedented precision.
\end{abstract}

\maketitle

\section{Introduction}

Over the past decades, scanning tunneling microscopy/spectroscopy (STM/STS) has evolved into one of the most versatile experimental techniques to explore the smallest length scales both topographically and spectroscopically \cite{phark_roadmap_2025}. One particular focus has been to move to lower temperatures and better energy resolution in order to gain access to the lowest energy scales in condensed matter physics \cite{assig_10_2013, schwenk_achieving_2020,  song_invited_2010,fernandez-lomana_millikelvin_2021, weerdenburg_scanning_2021, roychowdhury_30_2014, singh_construction_2013, matsui_development_2000}. These are, for example, related to spin phenomena, such as the hyperfine interaction \cite{willke_hyperfine_2018, kim_anisotropic_2022, stolte_single-shot_2025, farinacci_experimental_2022}, but also to superconductivity, such as the Josephson effect \cite{ast_sensing_2016,schwenk_achieving_2020,senkpiel_single_2020, naaman_fluctuation_2001}. 

The Josephson effect describes the tunneling of Cooper pairs between two superconductors and is one of the most impressive manifestations of quantum mechanics on macroscopic scales. It remains a cornerstone of condensed matter physics from macroscopic quantum tunneling to dynamical Coulomb blockade and constitutes the basic ingredient for superconducting quantum computing \cite{devoret_effect_1990,martinis_energy-level_1985,devoret_measurements_1985}. Moreover, Josephson junctions serve as exquisitely sensitive probes of quantum behavior at all energy scales. In the context of scanning tunneling microscopy (STM), the Josephson effect offers a unique window into the electromagnetic environment of the tunnel junction, making improvements in energy resolution not merely technical achievements but gateways to unexplored physics.

At low temperatures ($\lesssim 1$\,K), thermal fluctuations define no longer the principal energy scale $E_\text{th}=k_\text{B}T$ limiting the energy resolution in STM. Instead, electromagnetic fluctuations related to the charging energy $E_\text{C}=e^2/2C_\text{J}$ take over as the dominant energy scale. Here, $C_\text{J}$ is the junction capacitance, which places low temperature STMs in the dynamical Coulomb blockade (DCB) regime \cite{ingold_cooper-pair_1994,ingold_charge_1992}. This means that charge quantization effects emerge and an energy exchange of the tunneling electrons with the electromagnetic environment during tunneling has to be considered \cite{averin_incoherent_1990,devoret_effect_1990,ingold_charge_1992}. The $P(E)$-theory provides a framework for describing this energy exchange with the $P(E)$-function describing the probability of the environment to absorb/emit energy quanta from/to the tunneling Cooper pairs. The resulting spectral broadening constitutes an additional mechanism that ultimately limits the energy resolution in STS \cite{ast_sensing_2016}. 

This energy resolution limit applies for measurements along the voltage axis in STS. For the aforementioned spin phenomena, these limitations can be at least partially circumvented through the combination of STM with electron spin resonance spectroscopy (ESR-STM), which introduces the tunable microwave frequency as well as magnetic field sweeps as new energy axes. Accordingly, a much higher energy resolution reaching far into the neV regime can be achieved \cite{paul_generation_2016, drost_combining_2022, baumann_electron_2015}. Phenomena at other energy scales, such as the Josephson effect, have to be measured along the voltage axis, so that a continuous improvement of the energy resolution along the voltage axis remains desirable. Apart from sharper spectral features, improving the energy resolution always opens opportunities to access new energy scales that were hitherto impossible to see. In this regard, the Josephson effect in the DCB regime, which probes resonances in the electromagnetic environment \cite{jack_nanoscale_2015, jack_critical_2016}, carries the unique potential to resolve lower lying resonant features at longer wavelengths.

Here, we show that through the combination of local electromagnetic shielding and low-pass signal filtering directly at the low temperature STM scan head, we are able to greatly improve the spectroscopic energy resolution into the low $\upmu$eV range. We demonstrate this for two different STMs operating at 560\,mK and at 10\,mK base temperature, respectively. In both cases, the energy resolution improved by almost an order of magnitude. We attribute this improvement to an efficient shielding of the tunnel junction from high frequency radiation ($>1$\,MHz) as well as a capacitive shunting of the junction capacitance by the filter capacitors. As a result, we do not just find sharper spectral features in the measurements, but open the potential to combine low temperature STM with cavity quantum electrodynamics away from equilibrium (charge transfer). In this way, the Josephson current couples to the cavity resonances of the cylindrical scan head at centimeter length scales enabling cavity quantum electrodynamics with the STM \cite{blais_circuit_2021,wallraff_strong_2004}.

\section{Experimental Setup}

A schematic of the experimental setup is shown in \figref[a]{Fig1}. The trapezoid shape in the center represents the conically shaped scan head made of solid copper. Inside the scan head, there is the coarse motor (attocube $z$-stage), the tubular scan piezo, the tip and the sample \cite{assig_10_2013}. At the scan head, we have now installed two types of low-pass filters. There are $\pi$-filters \cite{pifilter}\ for the twisted pair cables connecting to the $xy$-scan piezo and the coarse motor as well as single line coaxial low-pass filters (Basel Precision Instruments, \cite{baselfilter,scheller_silver-epoxy_2014}) connecting to the current, bias voltage, and $z$-piezo. In this way, all electrical lines feeding into the scan head are filtered and the scan head provides a complete metallic enclosure. In addition, all lines (including electric connections that do not feed into the scan head, e.g.\ thermometry, heaters) that go into the cryostat are also filtered at the room temperature feedthroughs, which has been reported before \cite{assig_10_2013}. 

\begin{figure}
\centering
\includegraphics[width=\columnwidth]{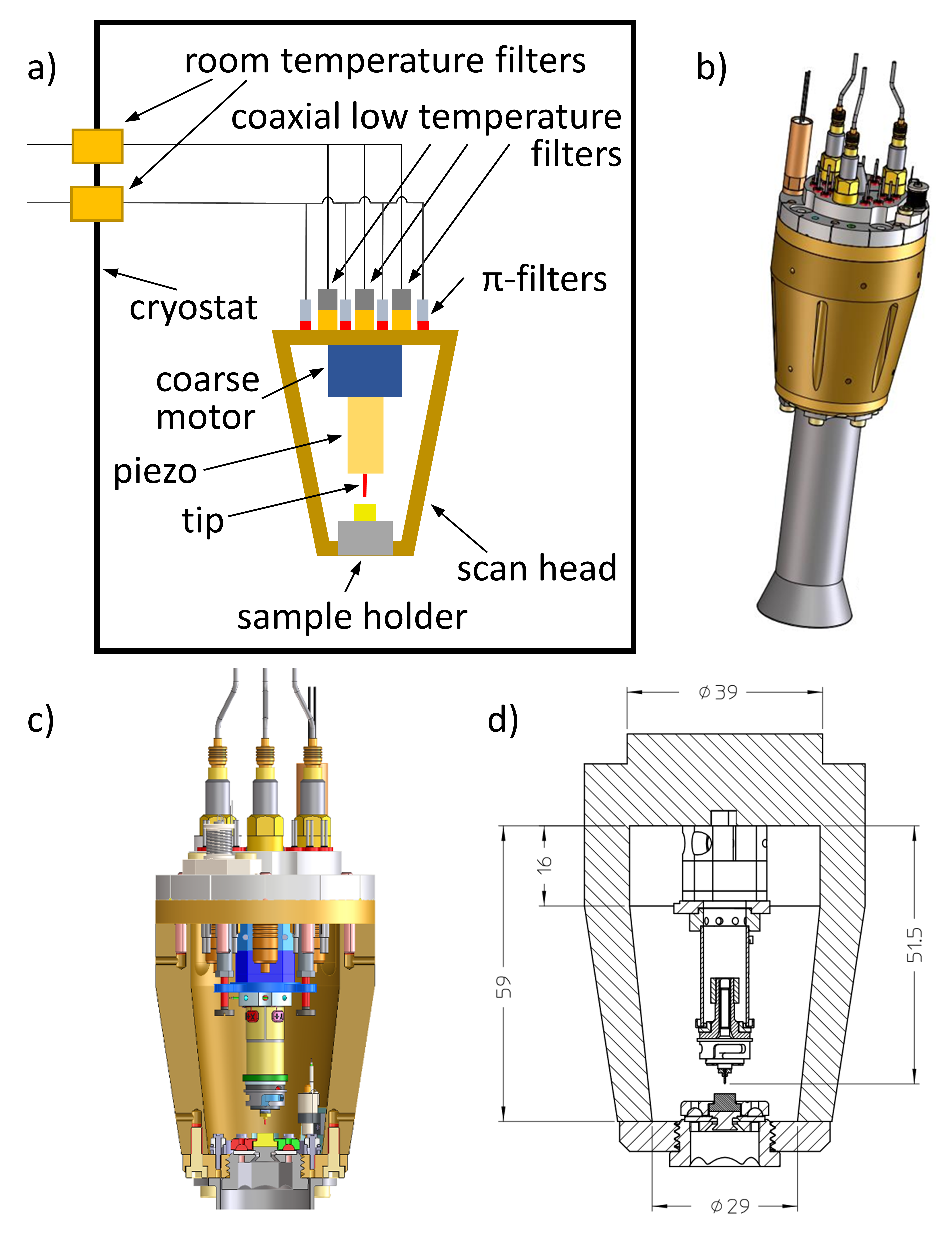}
\caption{\label{Fig1} \panelcaption{a} Schematic of the STM scan head (golden trapezoid) inside the cryostat (black rectangle) along with the positions of the low-pass filters. Every line that goes into the cryostat as well as into the scan head is filtered. \panelcaption{b} Three-dimensional drawing of the scan head (golden) along with the guidance cone (gray) on the bottom for tip and sample transfer and the electrical connections on the top along with the filters. \panelcaption{c} Cross sectional view of the three-dimensional drawing of the scan head. The coarse motor (attocube) is shown in blue with the scan piezo and the tip below. The sample holder (gray) is at the bottom along with the sample (yellow). \panelcaption{d} The two-dimensional drawing of the cross section in \panelsubcaption{c} along with the dimensions (in mm) describing the inside cavity of the scan head.}
\end{figure}

In \figref[b]{Fig1}, a drawing of the scan head can be seen with the three coaxial low-pass filters clearly visible on top and the guidance funnel on the bottom for tip and sample exchange. In \figref[c]{Fig1}, a cross section of the scan head can be seen showing the coarse motor, the scan piezo along with the tip, and the sample. The internal dimensions of the slightly conical scan head revealing the size of the cavity inside is shown in \figref[d]{Fig1}, with the dimensions given in millimeters. 

The measurements were done in two different STM setups, the mK-STM and the microwave STM (mw-STM). The mK-STM is a home-built setup operating at a base temperature of 10\,mK with a Janis dilution refrigerator \cite{assig_10_2013}. The mw-STM is also a home-built setup with a CryoVac Joule-Thompson refrigerator operating at a base temperature of 560\,mK. Both setups use the same scan head as shown in \figref{Fig1} with the only difference that the mw-STM has an additional port on the side of the conical wall with an antenna for radiating microwaves up to 100\,GHz into the tunnel junction \cite{drost_combining_2022,siebrecht_microwave_2023}. In addition, the mK-STM is located in the Precision Laboratory (PL) (low-noise environment, better grounding) and operated with a Nanonis RC5, while the mw-STM is located in a standard lab in the main institute building and operated with a Nanonis RC4. We note here that the choice of RC4 or RC5 did not change the energy resolution. 

All measurements reported here were done with a vanadium V(100) single crystal and a polycrystalline vanadium tip with a diameter of $0.25$\,mm. More details can be found in the Methods section. 

\section{Results and Discussion}

\sidecaptionvpos{figure}{c}
\begin{SCfigure*}
\centering\includegraphics[width=1.4\columnwidth]{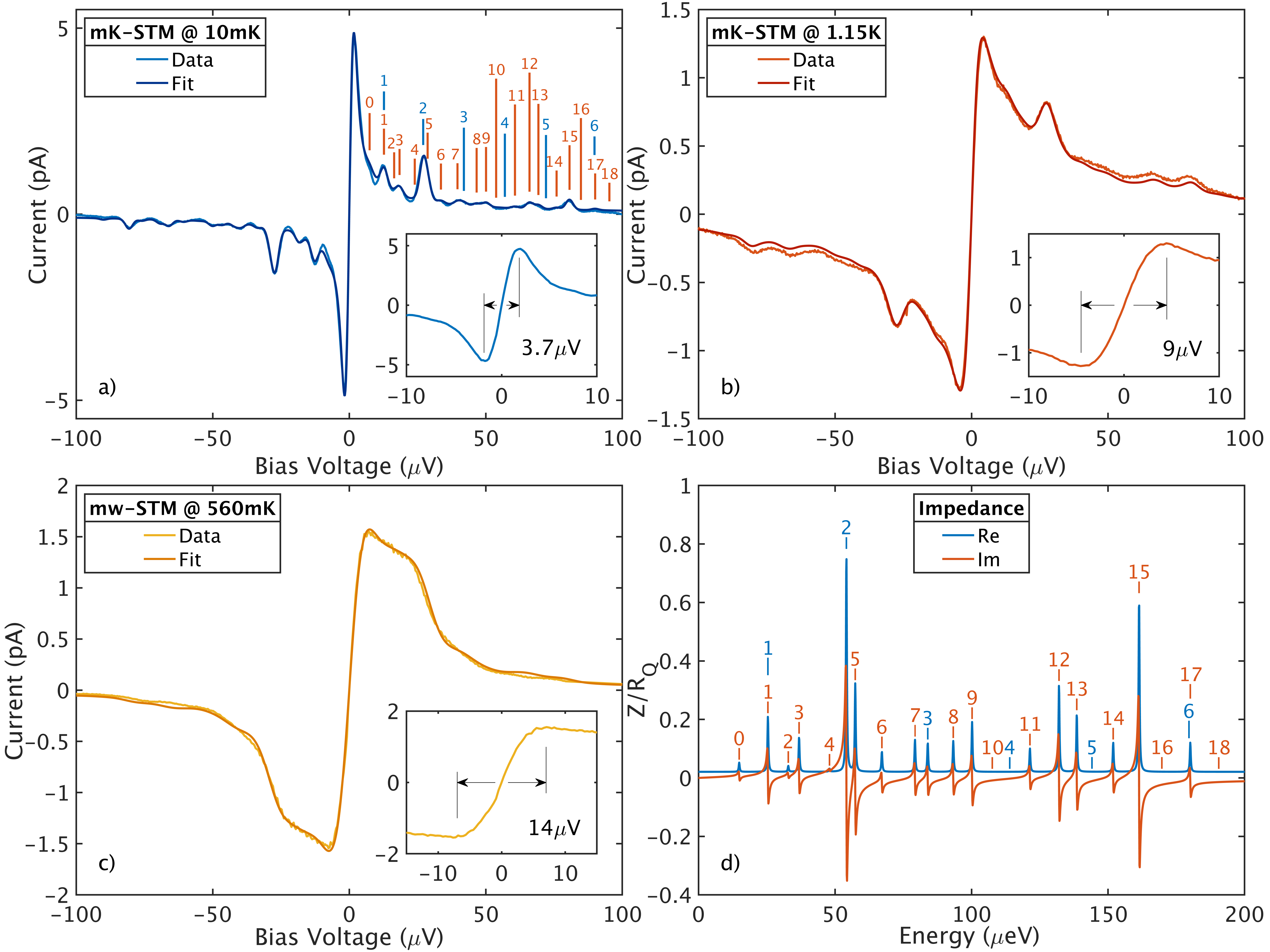}
\caption{\label{Fig2} \panelcaption{a} The measured Josephson current (light blue) as function of applied bias voltage in the mK-STM at 10\,mK along with the fit (dark blue). The vertical lines indicate the position of the resonances (blue: radial, red: axial). The resonances are clearly visible in the data. \panelcaption{b} Same measurement as in \panelsubcaption{a}, but at 1.15\,K. Due to the higher temperature, the resonances are washed out and cannot be individually identified anymore. \panelcaption{c} Same measurement as in \panelcaption{a} and \panelcaption{b}, but at the mw-STM and at 560\,mK. Even though the temperature is lower than in \panelcaption{b}, the stronger external broadening washes out the resonances more. The insets in \panelcaption{a} to \panelcaption{c} show a zoom in to zero bias showing the voltage difference between the positive and negative switching currents. \panelcaption{d} The real and imaginary parts of the environmental impedance are shown here. This is the impedance that was used to fit the data in \panelcaption{a} and \panelcaption{b} and with minor adjustments also in \panelcaption{c} (for details, see Table \ref{tab:params}). The setpoint current was 2\,nA at 4\,mV for all measurements.}
\end{SCfigure*}

\subsection{Benchmark Measurements}

The improvements in energy resolution can be most directly seen by measuring the Josephson current close to zero bias voltage. Its characteristic features react very sensitively to noise and changes in the environment, which can be used as a straightforward and model-free benchmark for the energy resolution \cite{schwenk_achieving_2020}. In \figref{Fig2}, the Josephson current within $\pm 100\,\upmu$V bias voltage is shown for the mK-STM at 10\,mK and 1.15\,K (\figref[a]{Fig2} and \figref[b]{Fig2}) and for the mw-STM at 560\,mK (\figref[c]{Fig2}). The characteristic antisymmetric shape with the highest Josephson current (switching current) close to zero bias voltage can be directly seen in each panel. At higher bias voltages, the peaks resulting from the interaction with environmental resonances due to DCB in each spectrum show distinct differences \cite{jack_nanoscale_2015,ast_sensing_2016,ingold_charge_1992}, which will be discussed below in more detail. Focusing on the peak close to zero bias voltage first, each panel has an inset with a zoom-in to about $\pm 10\,\upmu$V to enhance highest current peak. The benchmark value for the energy resolution is defined as the voltage difference between the switching currents at positive and negative bias voltage. This is indicated by the arrows in each inset along with the actual value.

The smallest benchmark value of 3.7\,$\upmu$V, we find for the mK-STM at 10\,mK (\figref[a]{Fig2}). This value is smaller than previously reported values of 10.6\,$\upmu$V at 10\,mK \cite{schwenk_achieving_2020} and $\approx 14\,\upmu$V at 80\,mK \cite{fernandez-lomana_millikelvin_2021}, where low-pass filtering at the low temperature STM scan head has been reported as well. The value of $\approx 14\,\upmu$V from \cite{fernandez-lomana_millikelvin_2021} is an extrapolated value to match with our definition of the benchmark (current maximum and minimum). Even though the filtering concept at the STM scan head is quite similar in all instances, our fully metallic scan head provides additional shielding of the tunnel junction from higher temperature radiation (see \figref{Fig1}), which could at least partially explain our improved energy resolution. We also want to point out that in these previous reports \cite{schwenk_achieving_2020,fernandez-lomana_millikelvin_2021} Al-Al junctions were used to measure the Josephson effect, whereas here, we use V-V junctions. This does not affect the outcome of the energy resolution measurement because the Josephson energy $E_\text{J}$ only scales the magnitude of the Josephson current, but does not change the shape of the spectrum. This is another advantage of this benchmark. Recent work has also highlighted the importance of accurately determining the junction temperature in mK-STM experiments using $P(E)$-theory \cite{temirov_determining_2023}, which corroborates our approach.

\begin{figure}
\centering\includegraphics[width=0.95\columnwidth]{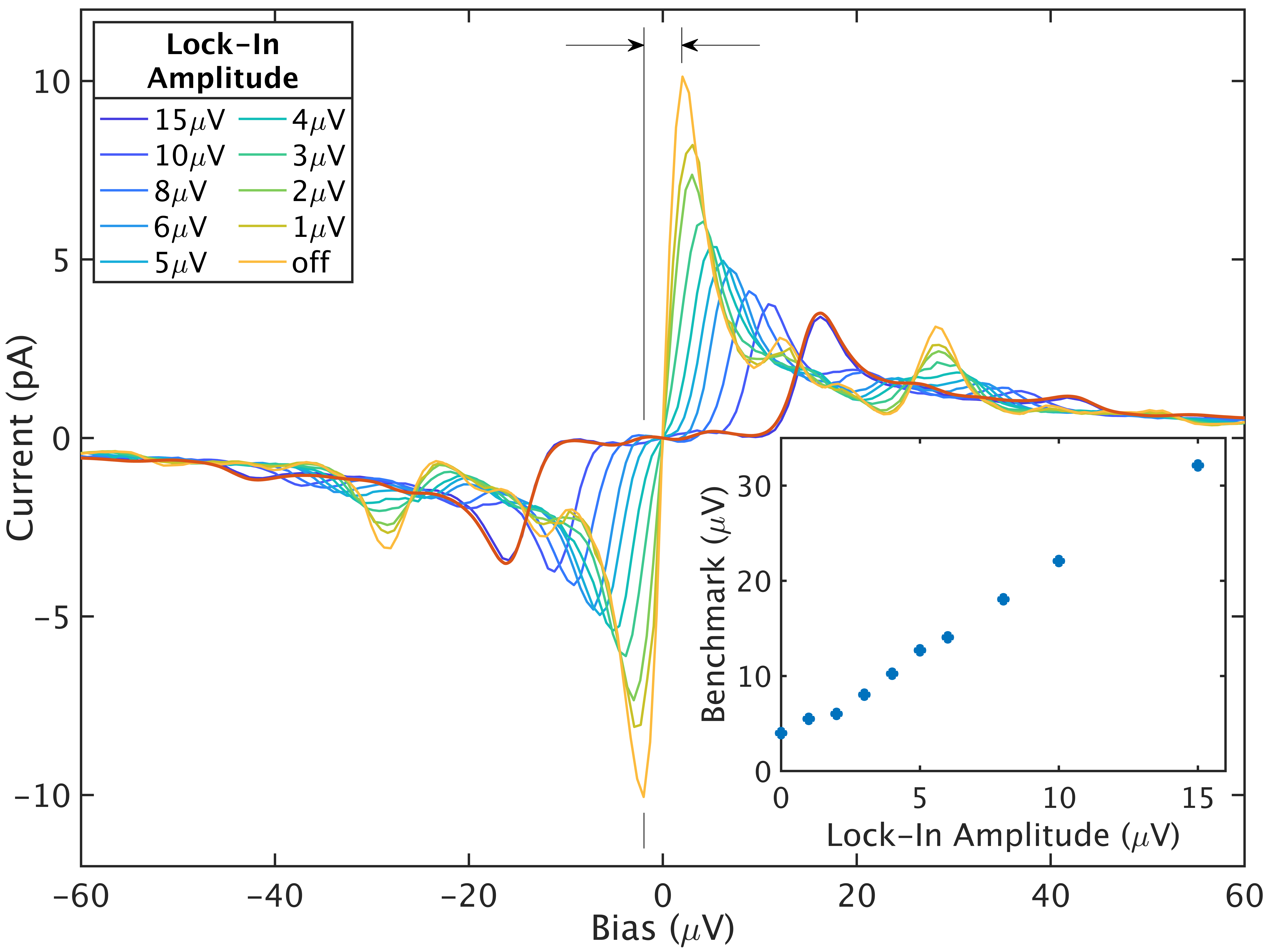}
\caption{\label{Fig6} The Josephson current is plotted as function of the applied bias voltage for different amplitudes of the lock-in amplifier. Even though the lock-in signal is not shown here, the broadening of the features due to the lock-in modulation is clearly visible. Already the smallest amplitude of 1\,$\upmu$V increases the benchmark value. The inset shows the benchmark value as function of the lock-in amplitude, which is the voltage difference between the positive and negative switching currents (see arrows). The red line is a convolution of the Josephson current without lock-in with a lock-in broadening of 15\,$\upmu$eV, which fits very well to the Josephson current that was measured with the same lock-in amplitude. The setpoint current was 3\,nA at 4\,mV for all measurements.} 
\end{figure}

Furthermore, unlike previous reports \cite{schwenk_achieving_2020,fernandez-lomana_millikelvin_2021}, we do not provide differential conductance measurements of the Josephson effect, because already the smallest lock-in amplitudes reduce the energy resolution, such that only the current measurement itself provides the least perturbed signal. This can be seen in \figref{Fig6}, where the Josephson current is shown as a function of different lock-in amplitudes. The Josephson current is already affected by lock-in amplitudes as low as 1\,$\upmu$V, such that for conductance measurements in general the effect of the lock-in amplifier on the energy resolution has to be taken into account. The red line in \figref{Fig6} shows a convolution of the Josephson current without lock-in with a lock-in broadening of 15\,$\upmu$eV, which fits very well to the Josephson current that was measured with the same lock-in amplitude. This demonstrates an overall consistency with the lock-in measurement and excludes other additional broadening in these measurements. The inset in \figref{Fig6} shows the increase of the benchmark value as a function of the lock-in amplitude.

Coming back to \figref[b]{Fig2}, the measurement at 1.15\,K was taken with the same V-V tunnel junction as in \figref[a]{Fig2} at 10\,mK. The principal current peaks are 9\,$\upmu$V apart (see inset). This higher benchmark value is expected from the higher temperature. In addition, the resonance peak structure at higher voltages looks much broader and less defined as in \figref[a]{Fig2}. We conclude that between 1.15\,K and 10\,mK, we observe a strong temperature dependence in the spectra, which we attribute to the improved energy resolution that makes these differences observable.

In \figref[c]{Fig2}, the Josephson current for the mw-STM at 560\,mK can be seen, which features a benchmark value of 14\,$\upmu$V (see inset). Even though the temperature is about half of the value than for the measurement in \figref[b]{Fig2}, the benchmark value is higher indicating a worse energy resolution. This can be explained by the general laboratory environment, in which the experiments are located (see Experimental Setup).  

\begin{table}[]
    \centering
    \begin{tabular}{r|r|r|r|r}
    Index $i$ $(n,l)$ & $A_i$ ($\upmu$eV) & $\omega_i$ ($\upmu$eV) & $E_{nl}$ ($\upmu$eV) & $\delta\omega (\%)$\\
    \hline\hline
      radial,1 (1,1) & 0.96 &  25.42 & 25.42 & 0.00\\
      2 (1,2) & (14.40) 7.20 &  54.16 &  54.16 & 0.00\\
      3 (1,3) & (4.80)  0.96 &  83.96 & 83.96 & 0.00\\
      4 (1,4) & 0.00 &  113.99 & 113.99 & 0.00\\ 
      5 (1,5) & 0.00 &  144.11 & 144.11 & 0.00\\ 
      6 (1,6) & 0.00 &  179.63 & 174.26 & 0.00\\
    \hline
    axial,0 (0,1) & (0.00) 0.30 &  14.91 & 14.91 & 0.00 \\
    1 (1,1) &  0.86 &  25.42 & 25.42 & 0.00\\  
    2 (2,1) &  0.19 &  32.88 & 31.27 & 5.17\\  
    3 (3,1) & (5.18) 1.15 &  36.83 & 39.11 & -5.81\\
    4 (4,1) & 0.10 &  47.98 & 47.98 & 0.00\\
    5 (5,1) & 2.88 &  57.41 & 57.41 & 0.00\\
    6 (6,1) & 0.67 &  67.16 & 67.16 & 0.00\\
    7 (7,1) & 1.06 &  78.28 & 77.1 & 2.82\\
    8 (8,1) & 1.01 &  93.30 & 87.19 & 7.01\\ 
    9 (9,1) & 1.63 &  100.20 & 97.36 & 2.92\\
    10 (10,1)& 0.00 &  107.59 & 107.59 & 0.00\\
    11 (11,1)& 0.77 &  121.38  & 117.87 & 2.98\\
    12 (12,1)& 2.88 &  132.03 & 128.19 & 2.99\\
    13 (13,1)& 1.92 &  138.54 & 138.54 & 0.00\\
    14 (14,1)& 0.96 &  151.88 & 148.91 & 1.99\\
    15 (15,1)& 5.76 & 161.35 & 159.30 & 1.29 \\
    16 (16,1)& 0.00 & 169.70 & 169.70 & 0.00 \\
    17 (17,1)& 0.96 & 180.12 & 180.12 & 0.00 
    \end{tabular}
    \caption{The parameters for the environmental resonances for the mK-STM are given in the table. For the index, $i$ refers to the experimental fit and the index in \figref[a]{Fig2} and $(n,l)$ refers to the modes in the cylindrical cavity model. Also, $A_i$ is the amplitude, $\omega_i$ is the fitted mode energy, $E_{nl}$ is the model mode energy, and $\delta\omega_i=(\omega_i-E_{nl})/E_{nl}$ is the deviation in mode energies. Values that differ for the mw-STM are given in parenthesis. All other resonance values are the same. Axial modes with indices higher than 17 have zero amplitude. The width of the Lorentzians $\lambda=0.2\,\upmu$eV is the same for all resonances. The other parameters for the mK-STM are $C_\text{J} = 1\,$fF, $R_\text{DC} = 269.29\,\upOmega$, as well as an additional Gaussian broadening with a full width half maximum of $\lambda_\text{ext}=3.0,\upmu$eV. The critical current is $I_\text{C}=552.05\,$pA and $I_\text{C}=589.05\,$pA for the data from the mK-STM and the mw-STM, respectively. The gap parameter for the vanadium sample is 750\,$\upmu$eV as well as for the vanadium tip it is 680\,$\upmu$eV and 750\,$\upmu$eV for the mK-STM and the mw-STM, respectively. The temperature is given in each panel of \figref{Fig2}., respectively. The Josephson energy $E_\text{J}$ can be calculated from the critical current $I_\text{C}$ as $E_\text{J} = \frac{\hbar}{2e} I_\text{C}$. For the mw-STM, the only other parameters that are different from the mK-STM are $C_\text{J} = 7.5\,$fF and  $\lambda_\text{ext}=10.0\,\upmu$eV. For the calculation of the modes, we use $R=20.5$mm, $L=59$mm, and $L^{\ast}=35mm$.}
    \label{tab:params}
\end{table}

\subsection{DCB Model Analysis}

We now go beyond the qualitative and model-free observations to a more quantitative analysis by applying the DCB model and $P(E)$-theory to describe the Josephson effect in the STM at very low temperatures \cite{devoret_effect_1990,averin_incoherent_1990,senkpiel_dynamical_2020}. In the DCB regime, the Josephson current is calculated from the $P(E)$-function as \cite{ast_sensing_2016,ingold_charge_1992,ingold_cooper-pair_1994,ingold_finite-temperature_1991}:
\begin{equation}
    I(V) = \frac{\pi e}{\hbar} |E_\text{J}|^2 \big(P(2eV)-P(-2eV)\big),
    \label{eq:iv}
\end{equation}
where $E_\text{J}$ is the Josephson energy, and $V$ is the applied bias voltage. The $P(E)$-function is the Fourier transform of the phase-phase correlation function $J(t)$ and captures the probability to exchange energy quanta (photons) with tunneling Coopers pairs. These functions are defined as
\begin{equation}
    P(E) = \int\limits^{\infty}_{-\infty}\frac{dt}{2\pi\hbar}e^{J(t) + i Et/\hbar}
\end{equation}
and
\begin{equation}
    J(t) = \int\limits^{\infty}_{-\infty}\frac{d\omega}{\omega}\frac{\text{Re}[Z_t(\omega)]}{R_\text{Q}}\frac{e^{i\omega t}-1}{1-e^{-\hbar\beta\omega}},
\end{equation}
where $\beta = 1/(k_\text{B} T)$ is the inverse temperature, $k_\text{B}$ is the Boltzmann constant, $T$ is the temperature, $\hbar$ is the reduced Planck constant, $e$ is the elementary charge, and $R_\text{Q}=h/(2e^2)$ is the resistance quantum. The total impedance $Z_t(\omega)$ depends on the environmental impedance as
\begin{equation}
    Z_t(\omega) = \frac{1}{i\omega C_\text{J} + Z^{-1}(\omega)},
    \label{eq:ztot}
\end{equation}
where $C_\text{J}$ is the junction capacitance and $Z(\omega)$ is the environmental impedance. In order to accomodate the many resonances, which appear in the Josephson current in \figref[a]{Fig2}, we adapt the environmental impedance as a sum of Lorentzians with a constant offset as
\begin{equation}
    Z(\omega) = R_\text{DC}\Big(1+\sum\limits_i \frac{A_i}{i(\omega\pm\omega_i) + \lambda}\Big),
\end{equation}
where $R_\text{DC}$ is the environmental dc resistance, $A_i$ is the resonance amplitude, $\omega_i$ is the resonance frequency, and $\lambda$ is the linewidth of the resonance, which we combine to one parameter for all resonances to keep the model simple (hence, no index $i$). To make the Lorentzians dimensionless, we use the amplitude, frequency as well as the linewidth in units of energy. 

We can fit this model to the experimental data at 10\,mK in \figref[a]{Fig2}, which is shown as the dark line. Apart from small details, we find excellent agreement with the experimental data. The values of the parameters used in the fit are given in Table \ref{tab:params}. To accommodate an unknown contribution from a general temperature independent noise broadening, we convolve the spectrum with a Gaussian having a full width at half maximum (FWHM) of $\lambda_\text{ext}$. For the fit of the data at 1.15\,K in \figref[b]{Fig2}, we use exactly the same parameters as in \figref[a]{Fig2} at 10\,mK except for the temperature value. Again, we find very good agreement with the experimental data. We note that the Gaussian noise broadening is temperature independent and therefore the same for both fits. This provides an overall consistency of the model and highlights the thermal evolution of the spectrum from 1\,K to 10\,mK. 

\begin{figure}
\centering\includegraphics[width=1\columnwidth]{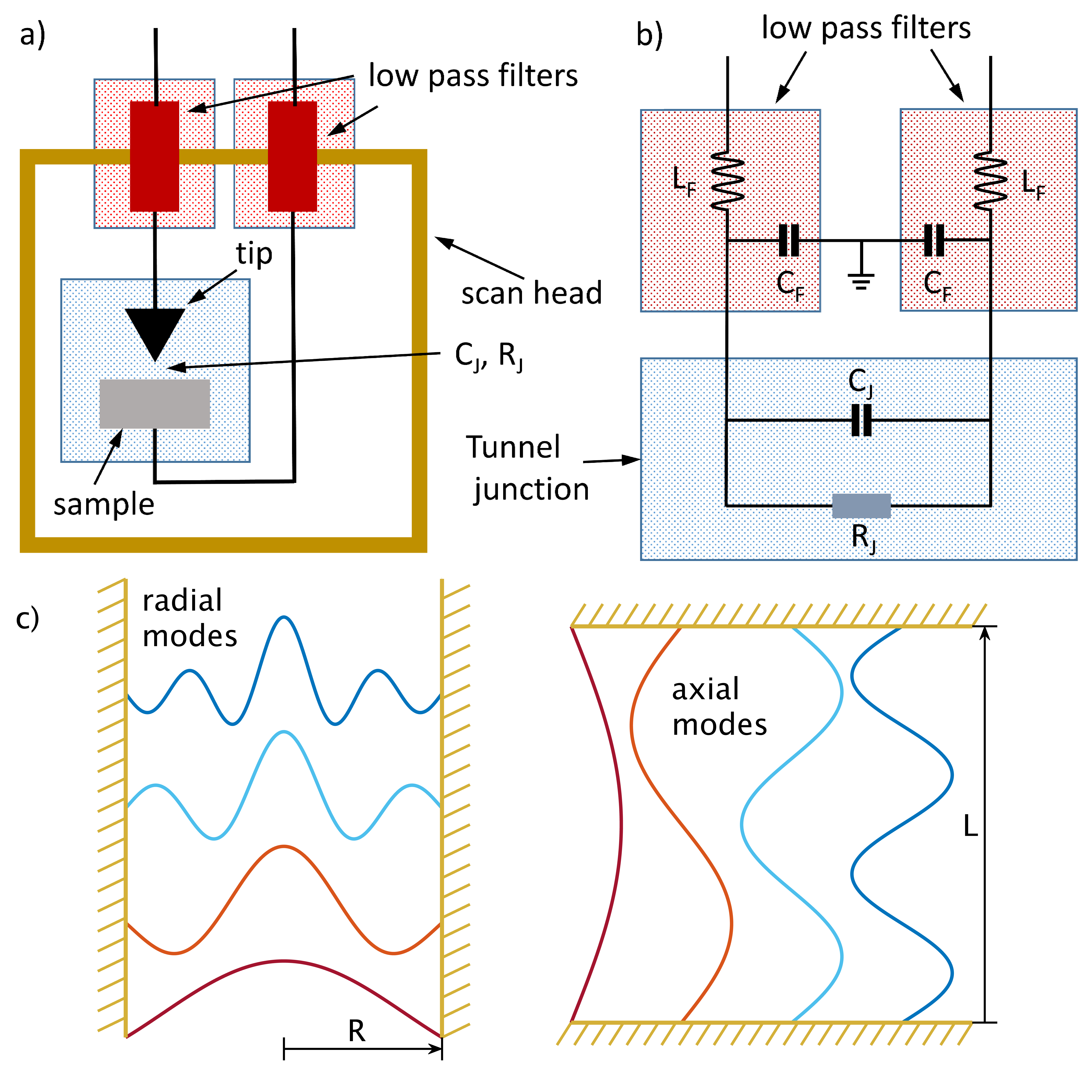}
\caption{\label{Fig5}\panelcaption{a} Schematic of the scan head with the tunnel junction (shaded blue) and the low-pass filters (shaded red). The tunnel junction is modelled by the junction capacitance $C_\text{J}$ and the tunneling resistance $R_\text{T}$. \panelcaption{b} Equivalent circuit diagram of the tunnel junction and the low-pass filters. The low-pass filters are modelled by the filter capacitance $C_\text{F}$ and the filter inductance $L_\text{F}$. The schematic makes it more clear how the junction capacitance is shunted by the filter capacitances. \panelcaption{c} Radial and axial components of the cylindrical cavity modes are shown. The first four modes are shown for each direction. The cavity is described by its radius $R$ and its length $L$.}
\end{figure}

Interestingly, in order to successfully fit the experimental data, we had to exclude the capacitive noise broadening that we have used before \cite{ast_sensing_2016}, which is temperature dependent. We attribute this to a shunting of the overall junction capacitance by the capacitances of the low temperature filters. This is conceptually shown in \figref[a]{Fig5}, where the scan head with the tunnel junction and the low-pass filters are shown. The equivalent circuit diagram is shown in \figref[b]{Fig5}, which shows more directly how the filter capacitors and the junction capacitance are connected in parallel. The effective junction capacitance $C_\text{J}$, which appears in the total environmental impedance should be seen as an effective coupling constant, which couples the tunneling electrons to the environmental resonances.  

The installation of the low-pass filters at the scan head has greatly improved energy resolution, which not only results in sharper spectral features, but reveals also the interaction of the tunneling Cooper pairs with the electromagnetic environment on a macroscopic scale. The environmental resonances visible in \figref[a]{Fig2} appear at much lower bias voltages than previously reported \cite{jack_nanoscale_2015}. Converting these bias voltages to energy, frequency, and wavelength of the corresponding electromagnetic radiation yields length scales on the order of millimeters to centimeters. These length scales correspond to the inner dimensions of the scan head, suggesting that its metallic enclosure acts as a microwave cavity. In the following, we show that the resonance structure in the Josephson spectrum can be attributed to the electromagnetic modes of the nearly cylindrical scan head.

\subsection{Cavity Mode Identification}

The environmental resonances, which can be seen in the spectrum in \figref[a]{Fig2} appear at much lower bias voltages than previously reported \cite{jack_nanoscale_2015}. In the spectrum in \figref[a]{Fig2}, there are no resonances probing the tip as an antenna also because we would expect them at higher bias voltages. This corroborates our previous statement that the energy resolution and the sensitivity of our setup has greatly increased and expands the parameter space for probing environmental resonances. Indeed, when converting the bias voltage to energy, frequency and wavelength of the corresponding electromagnetic radiation, we find length scales on the order of mm and cm. Hence, we have to look at these dimensions for identifying the origin of the electromagnetic resonances. 

Figure \ref{Fig1}(d) shows a cross section of the scan head with the inner dimensions of the metallic shell. We can approximate the conical shape of the scan head by a cylinder with radius $R$ and a finite length $L$. The electromagnetic modes that are found in a cylindrical cavity can be calculated using the Helmholtz equation (for details see the Supporting Information). These modes are schematically shown in \figref[c]{Fig5}. Here, we considered 100\% reflective walls, so that no attenuation occurs through losses due to imperfect reflection. The energies $E_{nl}$ of the resonance modes can be calculated from 
\begin{align}
E_{nl}= hf_{nl} = \hbar c\sqrt{\Big(\frac{x_l}{R}\Big)^2+\Big(n\frac{\pi}{L}\Big)^2},
\end{align}
where $n= 1,2,3,\ldots$ is the $n$-th axial mode and $x_l$ is the $l$-th zero of the Bessel function $J_0$ with $l = 1,2,3,\ldots$. In addition to these modes, there is one evanescent mode with the energy
\begin{align}
E_{0l}= hf_{0l} = \hbar c\sqrt{\Big(\frac{x_l}{R}\Big)^2-\Big(\frac{\pi}{L^\ast}\Big)^2},
\end{align}
where $L^\ast$ is an exponential decay length on the order of $L$. For the evanescent mode, the mode decays exponentially in the axial direction, which leads to a reduced mode energy compared to the minimal resonance mode $(n,l)=(1,1)$. The values for these parameters are given in the caption of Table \ref{tab:params}. We compare the fitted values of the mode energies with the calculated mode energies of the cylindrical cavity and find deviations within a few percent. These deviations can be understood from the geometry of the scan head, which is not cylindrical, but slightly conical (cf.\ \figref[d]{Fig1}) and also features parts inside (piezos, tip, and sample as well as an an antenna in the case of the mw-STM). The radius $R=20.5$\,mm used to calculate the modes is also slightly larger than the actual radius of the scan head (between 14.5\,mm and 19.5\,mm). We attribute this to imperfect reflection from the cylindrical walls, which are coated with a graphite layer, so that the radius $R$ appears larger because the modes are shifted to lower energies. In the vertical (axial) direction, the parallel plates are not coated with graphite and the length $L$ appears as in the real scan head. The width of each resonance directly determines the $Q$-factor of this respective mode. Using the definition $Q=\omega_i/2\gamma$, we find $Q$-factors between about 600 and 4500, which is reasonable considering that no optimization has been done.

We conclude that with this agreement between the fitted resonance energies and the calculated resonance energies of the cylindrical cavity, we are actually coupling to the cavity modes of our scan head during Josephson tunneling. With increasing voltage, whenever the resonance condition $E_{nl}=2 e V$ is met, Cooper pairs can more easily transfer their energy to the environment. The excess energy $2eV$ of a single Cooper pair is then deposited in the respective mode of the cavity. This opens up new possibilities in light-matter interaction, in particular cavity-quantum-electrodynamics and Josephson photonics \cite{hofheinz_bright_2011,westig_emission_2017,grimm_bright_2019,dambach_time-resolved_2015}. While the scan head is now optimized for tunneling experiments focusing on the tip and the sample, the scan head can be adapted to focus on light-matter interaction in the future. Here, a more reflective coating on the scan head walls to increase the $Q$-factor could be beneficial to induce a stronger interaction, possibly entering into a non-linear coupling regime. 

\begin{figure}
\centering\includegraphics[width=1\columnwidth]{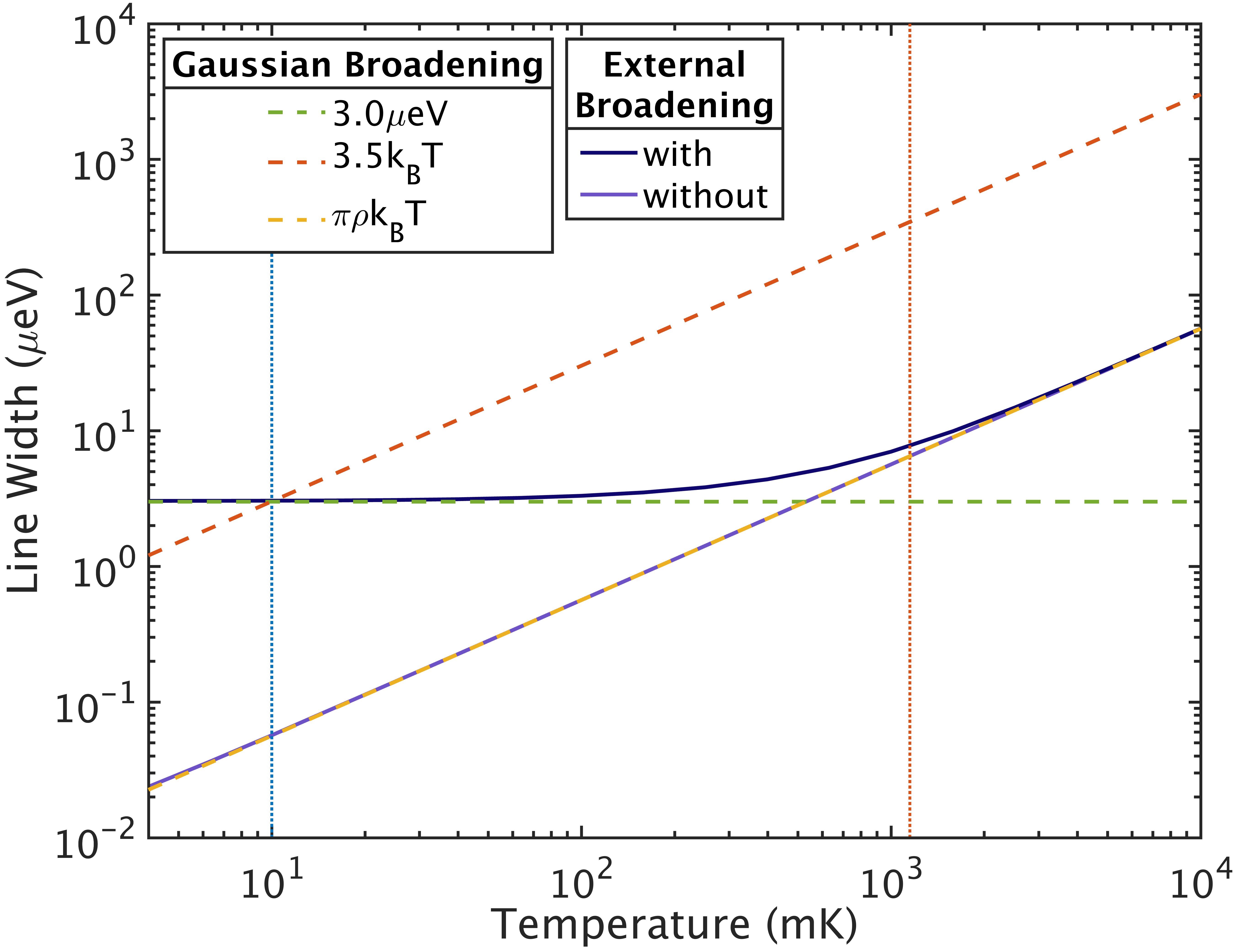}
\caption{\label{Fig4} The line width (full width at half maximum) is shown as function of temperature for thermal broadening, i.e.\ $3.5k_\text{B}T$, and $P(E)$-broadening assuming an ohmic environment. The $P(E)$-broadening is shown with (dark color) and without (light color) a constant contribution from an external noise source ($\lambda_\text{ext}=3\upmu$eV for the data of the mK-STM). With the external broadening, the $P(E)$-broadening converges to the constant value of the external broadening. Without the external broadening, the $P(E)$-broadening converges to $\pi\rho k_\text{B}T$. Since typically $\rho = \frac{R_\text{DC}}{R_\text{Q}} \ll 1$, the $P(E)$-broadening is much smaller than the thermal broadening (for details see text). The vertical dashed lines indicate the temperatures, at which the measurements were done with the mK-STM.}
\end{figure}

\subsection{Spectral Broadening Analysis}

We note that the benchmark values used in the Josephson measurement cannot be directly seen as an energy resolution in the sense of a spectral broadening. Still, the benchmark value gives a quick and model-independent measure of the energy resolution. A broadening in this sense comes from the $P(E)$-function, which can be seen as the energy resolution function in the STM because it enters into the quasiparticle tunneling current as an additional convolution \cite{ast_sensing_2016}. 

We can put the spectral broadening induced by the $P(E)$- function in relation to the temperature broadening due to the broadening of the Fermi-Dirac distribution. We compare the FWHM from thermal broadening and $P(E)$-broadening in \figref{Fig4} as a function of temperature. The thermal broadening from the finite width of the Fermi-Dirac distribution is $3.5k_\text{B}T$ (i.e.\ the FWHM of its derivative), which is shown as a red line in \figref{Fig4}. We note that in this work, there is no contribution from thermal broadening by the Fermi-Dirac distribution due to the presence of the superconducting gap in both tip and sample, which is the dominant energy scale here, i.e.\ $\Delta\gg k_\text{B}T$ \cite{rodrigo_use_2004,rodrigo_stm_2004}. The $P(E)$-function is slightly simplified by reducing the environmental impedance to an ohmic resistance. The difference to the actual environmental impedance discussed before is expected to be small since most of the modifications due to the cavity resonances occur at energies much larger than the FWHM. For the $P(E)$-function analyzed in \figref{Fig4}, we have used the fitting parameters from the mK-STM. In addition, we have included the constant (temperature independent) Gaussian broadening, which we had to include in the fit of our experimental data. This is shown in \figref{Fig4} for the darker line, which leads to a leveling at that constant value of $3\,\upmu$eV for low enough temperatures. The temperature values, at which the experimental data has been measured, are indicated as vertical dashed lines. This directly shows how an increasing temperature increases the broadening in the experimental data (from 10\,mK to 1.15\,K) and washes out the spectral features. We also find that for low temperatures the broadening due to the $P(E)$-function alone (i.e.\ no Gaussian broadening) converges to $\pi\rho k_\text{B} T$, where $\rho=R_\text{DC}/R_\text{Q}$ is the ratio of the environmental impedance at zero frequency to the quantum of resistance $R_\text{Q}=h/(2e^2)$. In a typical STM tunnel junction, $\rho$ is typically much smaller than one ($\rho\ll 1$). We find that if the capacitive noise broadening can be neglected, the $P(E)$-broadening is generally smaller than the thermal broadening (cf.\ Ref.\ \onlinecite{ast_sensing_2016}). This demonstrates that if the energy resolution is not limited by the thermal broadening due to the Fermi-Dirac distribution, e.g.\ by using superconducting tunnel junctions \cite{rodrigo_stm_2004,rodrigo_use_2004}, the energy resolution is limited by $P(E)$-broadening and other external contributions, which can be significantly lower than the thermal broadening (cf.\ \figref{Fig4}). This means that the energy resolution in the STM is, in principle, not limited by the thermal broadening due to the Fermi-Dirac distribution.

\section{Conclusion}

In summary, we have greatly enhanced the energy resolution in our scanning tunneling microscopes through internal filtering at base temperature and shielding at the scan head. We have demonstrated this through the benchmark value taken from the characteristic shape of the Josephson current in the STM. An effective shunting of the junction capacitance reduces the capacitive noise in the tunnel junction, which greatly reduces the spectral broadening from the $P(E)$-function. This enhanced energy resolution does not just allow us to observe sharper spectral features, but also opens up new possibilities for moving beyond existing limitations. We are now able to probe the electromagnetic modes of the scan head acting like a resonance cavity. This now opens fabulous perspectives to explore cavity quantum electrodynamics in the presence of charge transfer. Transferred charges drive cavity excitations which in turn give feedback on the Cooper pairs, thus giving rise, for example, to dressed states. In this way, we connect microscopic atomic scale tunneling with the macroscopic world on a centimeter scale. It also opens up new opportunities for measuring inelastic tunneling processes at extremely low energy scales with unprecedented resolution. 

\section{Methods}

The V(100) single crystal substrate was prepared through repeated cycles of Ar$^+$ sputtering, annealing to approximately 925\,K, and cooling to ambient temperature under ultra-high vacuum conditions to achieve an atomically flat surface. Surface reconstructions containing oxygen diffused from the bulk were observed. We generally did not observe a direct change in the superconductivity due to the presence of oxygen at the surface, consistent with previous reports \cite{koller_structure_2001,kralj_hraes_2003,huang_tunnelling_2020}. 

\section*{Acknowledgments}
We gratefully acknowledge stimulating discussions with Carlos Cuevas, Soumyaranjan Jhankar, Björn Kubala, Irena Padniuk, Ciprian Padurariu, and Dominik Zumbühl. This work was funded in part by the Center for Integrated Quantum Science and Technology (IQ$^\textrm{\small ST}$).

\clearpage
\newpage

\onecolumngrid
\begin{center}
\textbf{\large Supplementary Material}
\end{center}
\vspace{1cm}
\twocolumngrid

\setcounter{figure}{0}
\setcounter{table}{0}
\setcounter{equation}{0}
\renewcommand{\thefigure}{S\arabic{figure}}
\renewcommand{\thetable}{S\Roman{table}}
\renewcommand{\theequation}{S\arabic{equation}}

\section{Helmholtz Differential Equation with Dirichlet boundary conditions}
In order to determine the eigenmodes for a cylindrical cavity with finite length, we solve the Helmholtz differential equation with Dirichlet boundary conditions. Starting from the Maxwell equations\cite{jackson2012classical}, we have
\begin{align}
\nabla^2 E-\frac{n^2}{c^2} \frac{\partial^2 E}{\partial t^2} & =0 \\
\nabla^2 H-\frac{n^2}{c^2} \frac{\partial^2 H}{\partial t^2} & =0,
\end{align}
where $E$ is the electric field, $H$ is the magnetic field, $n$ is the refractive index, and $c$ is the speed of light. To solve these equations, we assume an electromagnetic waves with angular frequency $\omega$, which can be written as
\begin{align}
    & E(r, t)=E(r) e^{i \omega t} \\
    & H(r, t)=H(r) e^{i \omega t}
\end{align}
so that we can eliminate the explicit time dependence and we find
\begin{align}
&\nabla^2 E+k^2 E=0\\
&\nabla^2 H+k^2 H=0,
\end{align}
where we define the wave number $k=nk_0=\frac{n\omega}{c}=\frac{2 \pi n f}{c}=\frac{2n \pi}{\lambda}$ with $f$ being the frequency and $\lambda$ being the wavelength of the electromagnetic wave. 

These are the Helmholtz differential equations. To solve them in cylindrical coordinates, we write
\begin{align}
    \frac{1}{r} \frac{\partial}{\partial r}\left(r \frac{\partial F}{\partial r}\right)+\frac{1}{r^2} \frac{\partial^2 F}{\partial \theta^2}+\frac{\partial^2 F}{\partial z^2} +k^2 F=0
    \label{eq:Fcyl}
\end{align}
and use the separation of variables\cite{likharev2013essential} by writing
\begin{align}
    F(r, \theta, z)=R(r) \Theta(\theta) Z(z)
    \label{eq:Fdef}
\end{align}
to solve them. Inserting Eq.\ \eqref{eq:Fdef} into Eq.\ \eqref{eq:Fcyl}, we have
\begin{align}
\frac{d^2 R}{d r^2} \Theta Z+\frac{1}{r} \frac{d R}{d r} \Theta Z+\frac{1}{r^2} \frac{d^2 \Theta}{d \theta^2} R Z+\frac{d^2 Z}{d z^2} R \Theta+k^2 R \Theta Z=0 .
\end{align}

We multiply by $r^2 /(R \Theta Z)$, to separate the variables. We find
\begin{align}
    \left(\frac{r^2}{R} \frac{d^2 R}{d r^2}+\frac{r}{R} \frac{d R}{d r}\right)+\frac{1}{\Theta} \frac{d^2 \Theta}{d \theta^2}+\frac{r^2}{Z} \frac{d^2 Z}{d z^2}+k^2 r^2=0.
    \label{eq:RTZsep}
\end{align}

Since the solution must be periodic in $\theta$ by definition of the circular cylindrical coordinate system, the solution to the second term in Eq.\ \eqref{eq:RTZsep} must have a negative separation constant, which allows us to define the differential equation in $\Theta$ as
\begin{align}
    \frac{1}{\Theta} \frac{d^2 \Theta}{d \theta^2}=-m^2.
    \label{eq:DGTheta}
\end{align}
The solution to this equation is
\begin{align}
\Theta(\theta)=C_m \cos (m \theta)+D_m \sin (m \theta) .
\end{align}
Inserting Eq.\ \eqref{eq:DGTheta} into Eq.\ \eqref{eq:RTZsep}  and dividing by $r^2$ yields
\begin{align}
    \frac{1}{R} \frac{d^2 R}{d r^2}+\frac{1}{r R} \frac{d R}{d r}-\frac{m^2}{r^2}+\frac{1}{Z} \frac{d^2 Z}{d z^2}+k^2=0 .
    \label{eq:RZsep}
\end{align}
Because of the boundary conditions in the $z$-direction, we expect a periodic solution in $z$, so that the differential equation will have a negative separation constant, where $n$ is a real number. We write
\begin{align}
    \frac{1}{Z} \frac{d^2 Z}{d z^2}=-n^2,
    \label{eq:DGz}
\end{align}
which has the solution
\begin{align}
Z(z)=E_n \cos (n z)+F_n \sin (n z).
\end{align}

Inserting Eq.\ \eqref{eq:DGz} into Eq.\ \eqref{eq:RZsep} and multiplying by $R$ yields
\begin{align}
\frac{d^2 R}{d r^2}+\frac{1}{r} \frac{d R}{d r}+\left(k^2-n^2-\frac{m^2}{r^2}\right) R=0.
\end{align}
This is just a modified form of the Bessel differential equation\cite{arfken2011mathematical}, which has the solution
\begin{align}
R(r)=A_{m n} J_m\left(r \sqrt{k^2-n^2}\right)+B_{m n} Y_m\left(r \sqrt{k^2-n^2}\right),
\end{align}
where $J_n(x)$ and $Y_n(x)$ are Bessel functions of the first and second kind, respectively. The general solution is therefore
\begin{align}
& F(r, \theta, z)=\sum_{m,n}^{\infty} \left[A_{m n} J_m\left(r \sqrt{k^2-n^2}\right)+B_{m n} Y_m\left(r \sqrt{k^2-n^2}\right)\right] \nonumber\\
& \times\left[C_m \cos (m \theta)+D_m \sin (m \theta)\right]\left[E_n \cos (n z)+F_n \sin (n z)\right]
\end{align}

We now consider the scan head as a finite conducting cylinder with a unit source located at $(0,0,z_{0})$. The cylinder is bounded above and below by the planes $z=0$ and $z=L$ and on the sides by the radius $r=R$. Since we are only interested in the contribution which has a finite non-zero value at the point source, we have $B_{m n}=0$ and $m=0$. With the $z$ boundary conditions above, we also have $E_n=0$, so that we find the solution
\begin{align}
    F(r, z)=\sum_{n}^{\infty}A_{n} J_0\left(r \sqrt{k^2-k_n^2}\right) \sin (k_nz)
    \label{eq:F}
\end{align}
where 
\begin{align}
    k_n= n\frac{\pi}{L}, n=1,2,3,\ldots
\end{align}

For the radial boundary conditions, we let $k_lR$ be the $l$-th zero of $J_0$. The first few zeros $x_l$ of the Bessel function ($J_0(x_l) = 0$) are $x_l = \{$2.405, 5.520, 8.654, 11.792, 14.931, 18.071$\}$ for $l=1,\ldots, 6$\cite{arfken2011mathematical}. The boundary conditions only allow the following wave vector $k$
\begin{align}
k=\sqrt{k_l^2+k_n^2} = \sqrt{\Big(\frac{x_l}{R}\Big)^2+\Big(n\frac{\pi}{L}\Big)^2}.
\end{align}
Correspondingly, the resonance frequencies $f$ can be found at 
\begin{align}
f = \frac{ck}{2\pi} = \frac{c}{2\pi}\sqrt{\Big(\frac{x_l}{R}\Big)^2+\Big(n\frac{\pi}{L}\Big)^2}.
\end{align}

With a point source at $z=z_0$ and $r=0$, the homogeneous Helmholtz equation becomes inhomogeneous with the source term represented by a Dirac $\delta$-function. The modified equation in cylindrical coordinates becomes\cite{collin1990field}
\begin{align}
\nabla^2 F+k^2 F=-\frac{\delta(r) \delta\left(z-z_0\right)}{2 \pi r}
\end{align}
where the right-hand side represents the cylindrical coordinate form of the three dimensional Dirac $\delta$-function localized at $r=0, z=z_0$. To solve this equation we expand the $\delta$-function source in terms of the cavity's eigenfunctions
\begin{align}
\frac{\delta(r) \delta\left(z-z_0\right)}{2 \pi r}=\sum_{n,l}^{\infty}C_{n,l} J_0\left(k_lr \right) \sin (k_nz)
\end{align}
with $C_{n,l}$ as the coefficient for each term in the sum. 

Based on the following orthogonality conditions\cite{arfken2011mathematical}:

Angular part: For the trigonometric functions $\cos (m \theta)$ and $\sin (m \theta)$:
\begin{align}
\int_0^{2 \pi} \cos (m \theta) \cos \left(m^{\prime} \theta\right) d \theta=\pi \delta_{m m^{\prime}} \quad(m \neq 0)
\end{align}
with similar relations for $\sin (m \theta)$ and cross-terms vanishing.

Axial part: For $\sin \left(k_n z\right)$ with $k_n=n \pi / L$ :
\begin{align}
\int_0^L \sin \left(k_n z\right) \sin \left(k_{n^{\prime}} z\right) d z=\frac{L}{2} \delta_{n n^{\prime}}
\end{align}

Radial part: For Bessel functions $J_m\left(k_l r\right)$ with $k_l=x_l / R$, where $x_l$ is the $l$-th zero of $J_0$ :
\begin{align}
\int_0^R r J_m\left(k_l r\right) J_m\left(k_{l^{\prime}} r\right) d r=\frac{R^2}{2}\left[J_{m+1}\left(x_l\right)\right]^2 \delta_{l l^{\prime}}
\end{align}

Specifically, for $m=0$ :
\begin{align}
\int_0^R r J_0\left(k_l r\right) J_0\left(k_{l^{\prime}} r\right) d r=\frac{R^2}{2}\left[J_1\left(x_l\right)\right]^2 \delta_{l l^{\prime}}
\end{align}

The coefficients $C_{n,l}$ are determined via\cite{arfken2011mathematical}:
\begin{align}
C_{n,l}&=\frac{\sin\left(k_nz_0\right) J_0(0)}{2 \pi \int_0^R r J_0^2\left(k_lr \right) dr \int_{0}^{L} \sin^2 (k_nz) d z}\nonumber\\
&=\frac{\sin\left(k_nz_0\right) }{ \pi L \int_0^R r J_0^2\left(k_lr \right) d r}=\frac{2\sin\left(k_nz_0\right) }{\pi L R^2 J_1^2\left(k_lR \right)},
\end{align}
where the integrals normalize the eigenfunctions. The amplitude of each mode is determined by substituting the expanded source term into the inhomogeneous equation
\begin{align}
A_{n,l}(k)=\frac{C_{n,l}}{k^2-\left(k_l^2+k_n^2\right)} .
\end{align}
At resonance, when $k^2=k_l^2+k_n^2$, the amplitude diverges, corresponding to the cavity’s resonant frequencies. This is because in our boundary conditions, we have so far implied a perfect conductor resulting in perfect reflection of the wave. It is important to distinguish between the coefficients $A_n$ appearing in the homogeneous solution (Eq.\ \eqref{eq:F}) and the coefficients $A_{n,l}$ in the inhomogeneous solution. $A_n$ describes free oscillations of the cavity, while $A_{n,l}$ gives the response to an external drive with a point source. $A_n$ are constants for a given eigenmode, whereas $A_{n,l}$ depend on the driving frequency $k$ and diverge at resonance.

The solution $F(r,z)$ in Eq.~\eqref{eq:F} describes the homogeneous case with free oscillations of the cavity. When we include the external point source at $(0,0,z_0)$, we must solve the inhomogeneous equation. We denote this driven field by $\psi(r,z)$ to distinguish it from the free oscillation solution. The driven field is a superposition of all modes, each with the amplitude determined by the proximity to a resonance:
\begin{align}
\psi(r, z)=\sum_{n,l} \frac{C_{n,l} }{k^2-\left(k_l^2+k_n^2\right)} J_0\left(k_l r\right) \sin(k_nz).
\end{align}

In order to compare the amplitude for different resonant frequencies at the source point, we restrict ourselves to the $C_{n,l}$ coefficients. We find
\begin{align}
\psi_{n,l}(0, z_0)\propto\frac{\sin^2(k_n z_0)}{J_1^2\left(k_lR \right)} .
\end{align}

\section{Connection to Experimental Results}

The theoretical framework developed above allows us to identify the electromagnetic cavity modes of the cylindrical scan head. The result is that for a cylindrical cavity with radius $R$ and length $L$, resonances occur at frequencies determined by the zeros of the Bessel function $J_0$ in the radial direction and by integer multiples of $\pi/L$ in the axial direction. The amplitude at the source location (tunnel junction) is proportional to $\sin^2(k_n z_0)/J_1^2(k_lR)$, which determines the coupling strength between the Josephson current and each cavity mode. This explains why different modes appear with varying intensities in the experimental Josephson spectra shown in Fig.~2 of the main text. The excellent agreement between the calculated mode energies $E_{nl}$ and the fitted resonance positions $\omega_i$ (Table~I in the main text) validates this cavity model and demonstrates that the Josephson junction indeed couples to macroscopic electromagnetic modes spanning centimeter length scales. The small deviations (typically a few percent) can be attributed to the conical geometry of the actual scan head and the presence of internal components, which are not captured in the idealized cylindrical model. This theoretical treatment provides the foundation for understanding how atomic-scale quantum tunneling processes can probe and couple to the macroscopic electromagnetic environment.

\end{document}